\PassOptionsToPackage{hyphens}{url}
\documentclass[sn-mathphys]{sn-jnl}

\jyear{2022}
\title[Dedekind domains and class groups]{A formalization of Dedekind domains and class groups of global fields}

\author*[1]{\fnm{Anne} \sur{Baanen}}\email{t.baanen@vu.nl}
\author[2]{\fnm{Sander R.} \sur{Dahmen}}\email{s.r.dahmen@vu.nl}
\author[3]{\fnm{Ashvni} \spfx{Narayanan}}\email{a.narayanan20@imperial.ac.uk}
\author[4]{\fnm{Filippo A. E.} \sur{Nuccio Mortarino Majno di Capriglio}}\email{filippo.nuccio@univ-st-etienne.fr}

\affil[1]{\orgdiv{Department of Computer Science}, \orgname{Vrije Universiteit Amsterdam}, \orgaddress{\city{Amsterdam}, \country{The Netherlands}}}
\affil[2]{\orgdiv{Department of Mathematics}, \orgname{Vrije Universiteit Amsterdam}, \orgaddress{\city{Amsterdam}, \country{The Netherlands}}}
\affil[3]{\orgname{London School of Geometry and Number Theory}, \orgaddress{\city{London}, \country{United Kingdom}}}
\affil[4]{\orgname{Univ Lyon, Université Jean Monnet Saint-Étienne, CNRS UMR 5208, Institut Camille Jordan}, \postcode{F-42023} \city{Saint-\'Etienne}, \country{France}}

\usepackage{amsmath}
\usepackage{amsfonts}
\usepackage{amssymb}
\usepackage{url}
\usepackage{soul}
\usepackage{xcolor}
\usepackage{xspace}

\usepackage{listings}

\renewcommand{\C}{\mathbb{C}}
\newcommand{\inv}{\ensuremath{{}^{-1}}}
\newcommand{\lean}[1]{\texttt{#1}\xspace}
\newcommand*{\OK}[1][K]{\mathcal{O}_{#1}}
\newcommand*{\Cl}{\mathcal{C}\kern-.075em l}
\newcommand*{\Fq}[1][q]{\mathbb{F}_{#1}}
\newcommand*{\refFLT}[1]{($\mathrm{FLT}_{#1}$)}
\DeclareMathOperator{\Tr}{Tr}
\newcommand{\mathlib}{\textsf{mathlib}\xspace}
\newcommand{\N}{\mathbb{N}}
\renewcommand{\R}{\mathbb{R}}

\newcommand{\QQ}{\mathbb{Q}}
\renewcommand{\Z}{\mathbb{Z}}
\DeclareMathOperator{\Frac}{Frac}
\DeclareMathOperator{\card}{card}
\DeclareMathOperator{\Mod}{mod}

\definecolor{keywordcolor}{rgb}{0.7, 0.1, 0.1}   
\definecolor{commentcolor}{rgb}{0.4, 0.4, 0.4}   
\definecolor{symbolcolor}{rgb}{0.4, 0.4, 0.4}    
\definecolor{sortcolor}{rgb}{0.1, 0.5, 0.1}      

\DeclareUnicodeCharacter{02E3}{\ensuremath{{}^\times}}
\DeclareUnicodeCharacter{03B1}{\ensuremath{\alpha}}
\DeclareUnicodeCharacter{03C3}{\ensuremath{\sigma}}
\DeclareUnicodeCharacter{2081}{\ensuremath{_1}}
\DeclareUnicodeCharacter{2082}{\ensuremath{_2}}
\DeclareUnicodeCharacter{2090}{\ensuremath{_a}}
\DeclareUnicodeCharacter{2097}{\ensuremath{_l}}
\DeclareUnicodeCharacter{2115}{\ensuremath{\N}}
\DeclareUnicodeCharacter{211A}{\ensuremath{\QQ}}
\DeclareUnicodeCharacter{2124}{\ensuremath{\Z}}
\DeclareUnicodeCharacter{2211}{\ensuremath{\sum}}
\DeclareUnicodeCharacter{2223}{\ensuremath{\mid}}
\DeclareUnicodeCharacter{2264}{\ensuremath{\l}}
\DeclareUnicodeCharacter{2286}{\ensuremath{\subseteq}}
\DeclareUnicodeCharacter{22A4}{\ensuremath{\top}}
\DeclareUnicodeCharacter{22A5}{\ensuremath{\bot}}
\DeclareUnicodeCharacter{29F8}{\ensuremath{\textfractionsolidus}}

\begin{document}

\abstract{
Dedekind domains and their class groups are notions in commutative algebra that are essential in algebraic number theory.
We formalized these structures and several fundamental properties, including number theoretic finiteness results for class groups, in the Lean prover as part of the \mathlib mathematical library.
This paper describes the formalization process, noting the idioms we found useful in our development and \mathlib's decentralized collaboration processes involved in this project.

}

\maketitle

\section{Introduction}

In its basic form, number theory studies properties of the integers $\Z$
and its fraction field, the rational numbers $\QQ$.
For the sake of generalization, as well as for providing powerful techniques to answer questions about the original objects $\Z$ and $\QQ$,
it is worthwhile to study finite extensions of $\QQ$, called \emph{number fields}, as well as their \emph{rings of integers} (Section~\ref{sec math background}),
whose relations mirror the way $\QQ$ contains $\Z$ as a subring.
In this paper, we describe our project aiming at formalizing these notions and some of their important properties. Our goal is not to get to the definitions and properties as quickly as possible; rather, we lay the foundations for future work, as part of a natural and more general theory.

In particular, our project resulted in formalized definitions and elementary properties of
number fields and their rings of integers (Section~\ref{sec:ring-of-integers}),
Dedekind domains (Section~\ref{sec:Dedekind-domain}),
and the ideal class group and class number (Section~\ref{sec:class-number}).
Apart from the very basics concerning number fields, these concepts were not formalized before as far as we know.
We note that a formal definition of the class number is an essential requirement for the use of theorem provers in modern number theory research.
The main proofs that we formalized show
that two definitions of Dedekind domains are equivalent (Section \ref{sec:equivalence}),
that the ring of integers of a number field is a Dedekind domain (Section \ref{sec:integral-closure})
and that the class group of the ring of integers of a number field is finite (Section \ref{sec:class-number}).
In fact, most of our results for number fields are also obtained in the more general setting of \emph{global fields}.

Our work is developed as part of the mathematical library \mathlib~\cite{mathlib} for the Lean 3 theorem prover~\cite{lean-prover}.
The formal system of Lean is a dependent type theory augmented with quotient types and classical reasoning, both of which are commonly used in \mathlib~(Section~\ref{sec:lean-mathlib}).
As we finished parts of our work, we contributed these to \mathlib.
We, in turn, used results contributed by others after the start of the project.
At several points, we had just merged a formalization into \mathlib that another contributor needed,
immediately before they contributed a result that we needed.
Due to the decentralized organization and fluid nature of contributions to \mathlib, its contents are built up of many different contributions from over 250 different authors.
Attributing each formalization to a single set of main authors would not do justice to all others whose additions and tweaks are essential to its current use.
Therefore, we will make clear whether a contribution is part of our project or not, but we will not stress who we consider to be the main authors.

The source files of the formalization have been contributed to \mathlib.
We have preserved the development branch that this paper is based on\footnote{\url{https://github.com/leanprover-community/mathlib/tree/dedekind-domain-dev}}.
We also maintain a
repository%
\footnote{\url{https://github.com/lean-forward/class-number-journal}}
containing the source code referred to in this paper.

This paper is an extended version of a paper published in the ITP 2021 conference proceedings~\cite{ClassGroupsITP2021}.
The additions to this paper, apart from several clarifications and enhancements throughout the text, mainly concern the following.
\begin{itemize}
 \item Code samples throughout have been updated to reflect parts of our formalization contributed to \mathlib after the previous publication and to incorporate changes in \mathlib after contribution.
 \item Instead of only considering class groups of Dedekind domains, we briefly describe class groups for general domains; see the end of Section~\ref{sec math background} and Section~\ref{subsec:class_group}.
 \item The new Section~\ref{sec:lean-mathlib} gives a more detailed explanation of Lean as used in \mathlib, in particular the use of typeclasses and bundling.
 \item We discuss \emph{definitional equality} in Lean in the context of overlapping typeclass instances in Section~\ref{sec:field_extension}.
 \item The further evolution of fraction rings in \mathlib is discussed at the end of Section~\ref{subsection : fields of fractions}.
 \item We elaborate on invertibility and unique factorization of ideals in Dedekind domains in Sections~\ref{subsection:frac_ideals} and \ref{subsec:unique_ideal_factorization}.
 \item We give more details on the proof of finiteness of the class group and its formalization in Section~\ref{subsec:finiteness}.
 \item We elaborate on future directions in Section~\ref{sec:future_directions}, including research relying on the formalization described here.
\end{itemize}

\section{Mathematical background}\label{sec math background}

Let us now introduce some of the main objects we study, described informally. We assume some familiarity with basic ring and field theory.

A \emph{number field} $K$ is a finite extension of the field $\QQ$, and as such has the structure of a finite dimensional vector space over $\QQ$; its dimension is called the \emph{degree} of $K$.
The easiest example is $\QQ$ itself, and the two-dimensional cases are given by the quadratic number fields
$\QQ(\sqrt{d})=\{a+b\sqrt{d} : a,b \in \QQ\}$
where $d \in \Z$ is not a square. Similarly, adding a cubic root $\sqrt[3]{d}$ of some $d\in\Z$ which is not a cube leads to the number field $\QQ(\sqrt[3]{d})$: it has degree $3$ but not all cubic number fields arise in this way.
An example of a cubic number field that is not of this form, and that will occupy us later for other interesting features, is $\QQ(\alpha_0)=\{a+b\alpha_0+c \alpha_0^2: a,b,c \in \QQ\}$, where $\alpha_0$ is the unique real number satisfying $\alpha_0^3 + \alpha_0^2 - 2\alpha_0 + 8=0$.
In general, taking any root $\alpha$ of an irreducible polynomial of degree $n$ over $\QQ$ yields a number field of degree~$n$, namely
$\QQ(\alpha)=\{c_0+c_1\alpha+\ldots+c_{n-1} \alpha^{n-1} : c_0,c_1,\ldots,c_{n-1} \in \QQ \}$,
and, up to isomorphism, all number fields of degree $n$ arise in this way.

The \emph{ring of integers} $\OK$ of a number field $K$ is defined as the integral closure of $\Z$ in $K$, namely
\[
  \OK := \left\{x \in K : f(x)=0 \text{ where $f \in \Z[x]$ is a monic polynomial}\right\},\]
where we recall that a polynomial is called \emph{monic} if its leading coefficient equals~$1$.
While it might not be immediately obvious that $\OK$ is a ring, this follows from general algebraic properties of integral closures.
Some examples of rings of integers are the following. Taking $K=\QQ$, we get $\OK=\Z$ back.
For $K=\QQ(i)=\QQ(\sqrt{-1})$ we get that $\OK$ is the ring of Gaussian integers $\Z[i]=\{a+bi : a,b \in \Z\}$.
But for $K=\QQ(\sqrt{5})$ we do \emph{not} simply get $\Z[\sqrt{5}]=\{a+b\sqrt{5} : a,b \in \Z\}$ as $\OK$, since the golden ratio $\varphi:=(1+\sqrt{5})/2\not\in \Z[\sqrt{5}]$ satisfies the monic polynomial equation $\varphi^2-\varphi-1=0$; hence by definition, $\varphi \in \OK$.
It turns out that $\OK=\Z[\varphi]=\{a+b\varphi : a,b \in \Z\}$.
Finally, if $K=\QQ(\alpha_0)$ with $\alpha_0$ as before, then $\OK=\{a+b \alpha_0+c (\alpha_0+\alpha_0^2)/2 : a,b,c \in \Z\}$, illustrating that explicitly writing down $\OK$ can quickly become complicated.
Further well-known rings of integers are
the Eisenstein integers $\Z[(1+\sqrt{-3})/2]$ and the ring $\Z[\sqrt{2}]$.

Thinking of $\OK$ as a generalization of $\Z$, it is natural to ask which of its properties
still hold in $\OK$ and, when this fails, if a reasonable weakening does.

An important property of $\Z$ is that it is a principal ideal domain (PID), meaning that every ideal is generated by one element. This implies that every nonzero nonunit element can be written as a
finite product of prime elements, which is unique up to reordering and multiplying by $\pm 1$.
A ring where this holds is called a unique factorization domain, or UFD.
For example, $6$ can be factored in primes in $4$ equivalent ways, namely $6=2\cdot 3=3\cdot2=(-2)\cdot (-3)=(-3) \cdot (-2)$.
In fact, the previously mentioned examples of rings of integers are UFDs, but this is certainly not true for all rings of integers. For example, unique factorization \emph{does not} hold in $\mathcal{O}_{\QQ(\sqrt{-5})}=\Z[\sqrt{-5}]$
: it is easy to prove that $6=2\cdot3$ and $6=(1+\sqrt{-5}) (1-\sqrt{-5})$ provide two essentially different\footnote{By ``essentially different'' we mean that one factorization cannot be obtained from the other via multiplication by units.} ways to factor $6$ into prime elements of $\Z[\sqrt{-5}]$.

As it turns out, there is a way to weaken this notion of unique factorization in a meaningful way. Namely, by considering factorization of \emph{ideals} instead of elements; given a number field $K$, with ring of integers $\OK$, a beautiful and classical result by Dedekind shows that every nonzero ideal of $\OK$ can be factored as a product of prime ideals in a unique way, up to reordering.

Although unique factorization in terms of ideals is of great importance, it is still interesting, and sometimes
necessary, to also consider factorization properties in terms of elements.
We have already mentioned that unique factorization in $\Z$ follows from the fact that every ideal is generated by a single element.
Now, it is convenient to extend the notion of ideals of $\Z$ to that of $\emph{fractional ideals}$. 
These are additive subgroups of $\QQ$ of the form $\frac{1}{d} I$ with $I$ an ideal of $\Z$ and $d$ a nonzero integer.
When the distinction is important, we refer to an ideal $I \subseteq \Z$ as an \emph{integral ideal}.
The nonzero fractional ideals of $\Z$ naturally form a multiplicative group (whereas, for instance, there is no integral ideal $I\subseteq \Z$ such that $I*(2\Z)=(1)$).
The statement that every ideal is generated by a single element
translates to the fact that the quotient group of nonzero fractional ideals modulo $\QQ^\times$ is trivial (where $\frac{a}{b} \in \QQ^\times$ corresponds to~$\frac{1}{b} a \Z$, and the multiplicative group of invertible elements of a ring $R$ is denoted by $R^{\times}$).

It turns out that this quotient group can be defined for every ring of integers~$\OK$.
The fundamental theoretical notion beneath this construction is that of Dedekind domains: these are integral domains $D$ which are Noetherian (every ideal of $D$ is finitely generated), integrally closed (if an element $x$ in the fraction field $\Frac D$ of $D$ is a root of a monic polynomial with coefficients in $D$, then actually $x \in D$), and of Krull dimension at most $1$ (every nonzero prime ideal of $D$ is maximal).
It can be proved that the nonzero fractional ideals of a Dedekind domain~$D$
form a group under multiplication, and that the quotient of this group by the image of the natural embedding of $(\Frac D)^\times$ is called the \emph{\textup{(}ideal\textup{)} class group} $\Cl_{D}$.
For later reference, fractional ideals generated by one element of $\Frac D$ are called principal fractional ideals, so the image of the natural embedding of $(\Frac D)^\times$ consists exactly of the nonzero principal fractional ideals.

What is arithmetically crucial is the theorem ensuring that the ring of integers $\OK$ of every number field $K$ is a Dedekind domain,
and that in this case the class group $\Cl_{\OK}$ is actually \emph{finite}.
In particular, $\Cl_{\OK}$ can be seen as ``measuring'' how far ideals of $\OK$ are from being generated by a single element and,
consequently, as a measure of the failure of unique factorization.
The order of $\Cl_{\OK}$ is called the \emph{class number} of $K$.
Intuitively, then, the smaller the class number, the fewer factorizations are possible.
In particular, the class number of $K$ is equal to $1$ if and only if $\OK$ is a UFD.

The statements in the previous paragraph also hold for \emph{function fields}, namely fields which are finite extensions of $\Fq(t) = \Frac \Fq[q][t]$, where $\Fq[q][t]$ stands for the ring of univariate polynomials (in a free variable $t$) with coefficients in a finite field $\Fq$ with $q$ elements. Recall that when $q$ is a prime number, $\Fq$ is simply the field $\Z/q\Z$.
A field which is either a number field or a function field is called a \emph{global field}.

The concept of class group actually not only makes sense for Dedekind domains but more generally for (at least) any integral domain $R$ as follows. While the nonzero fractional ideals of $R$ in general need not be a group, they do form a commutative monoid. Hence, the invertible fractional ideals of $R$ form a group, and the class group of $R$ (denoted $\Cl_{R}$) is now defined as the quotient of this group by the image of the natural embedding of $(\Frac R)^\times$.

In upcoming sections we will describe the formalization of the above concepts as part of \mathlib.

\section{Lean and \mathlib} \label{sec:lean-mathlib}

The formal system of Lean is a dependent type theory based on the calculus of inductive constructions.
This means that each element \lean{e} has a unique type \lean{t}, written \lean{e : t}.
The natural number 0 has type \lean{ℕ}, and the rational 0 has type \lean{ℚ}.
One can then identify \lean{0 : ℕ} with \lean{0 : ℚ} using a map \lean{ℕ → ℚ} called a coercion (denoted by the arrow \lean{↑} or left implicit);
that is, \lean{(0 : ℚ) = ↑(0 : ℕ)}.
Types have types too, for example \lean{ℕ : Type}.
The full hierarchy consists of an impredicative universe \lean{Prop} sitting at the bottom of a noncumulative chain \lean{Prop $:$ Type $:$ Type 1 $:$ \mbox{Type 2} $:$ \dots}\,;
``an arbitrary \lean{Type u}'' is abbreviated as \lean{Type*}.
Propositions correspond to elements of \lean{Prop}, while a (verified) proof of the proposition \lean{P : Prop} corresponds to an element \mbox{\lean{p : P}}.
In addition to these features commonly found in a dependent type theory, Lean provides proof irrelevance, quotient types and classical reasoning.
Proof irrelevance means that for any proposition \lean{P : Prop}, any two proofs \mbox{\lean{p₁ p₂ : P}} are judged equal by the system.
These features are all commonly used in \mathlib.

Lean uses typeclass inference to automatically infer properties of certain objects.
If we define a structure with the keyword \lean{class},
then one can supply values for the class that Lean will automatically infer, by tagging these with \lean{instance}.
As an example, consider a ring $R$ with a subring $S$.
The instance \lean{subring.to\_ring} says that $S$ is also a ring.
Consequently, one can now use lemmas about rings for $S$ without having to invoke \lean{subring.to\_ring}.
We put the implicit arguments to be inferred by the typeclass system in square brackets.
Other implicit arguments remain in curly brackets, while explicit arguments go in round brackets. As an example, consider:
\begin{lstlisting}
theorem pow_succ {M : Type u} [monoid M] (a : M) (n : ℕ) :
  a ^ (n + 1) = a * a ^ n
\end{lstlisting}
When invoking this theorem, one must provide the explicit arguments \lean{a}, which has type \lean{M}, and a natural number \lean{n}.
As a result, Lean can determine the value of \lean{M} through \emph{unification} and can then use the typeclass system to infer a value for \lean{[monoid M]}.

The flagship general-purpose mathematical library for Lean is \mathlib; other libraries are available for more specialized purposes.
Organizationally, \mathlib is characterized by a distributed and decentralized community of contributors, a willingness to refactor its basic definitions, and a preference for small, yet complete, contributions over larger projects added all at once.
In this project, as part of the development of \mathlib, we followed this philosophy by contributing pieces of our work as they were finished.
In turn, we used other \mathlib contributors' results as they were made available.

There is a variety of tactics available in \mathlib such as
\lean{simp} (simplifies the main goal target using lemmas tagged with the attribute \lean{[simp]}),
\lean{library\_search} (tries to close the current goal by applying a lemma from the library) and
\lean{ring} (proves equality of polynomial expressions over commutative (semi)rings).
Lean uses these to simplify the statement or to close the goal.
These are very efficient when working with proofs that are calculation heavy,
or that follow from a small number of easy (or mathematically trivial) steps.

\subsection{Use of typeclasses and bundling} \label{sec:typeclasses}
Typeclasses were originally introduced in Haskell as a mechanism for operator overloading~\cite{typeclasses-haskell},
and are used throughout Lean's core library and \mathlib to endow types with mathematical structures consisting of both operators and their properties~\cite{mathlib}.
When the elaborator sees a function with an instance parameter being applied, such as the \lean{[monoid M]} parameter of \lean{pow\_succ a n}, a Prolog-like search is started to automatically synthesize a suitable value for this parameter.
Each of the local parameters and the declarations marked as \lean{instance} is tested in turn to see if their type matches the expected type of the instance synthesis.
All instance parameters of candidate instances are themselves recursively inferred, until either a suitable term is constructed or no more candidates remain; an error is raised in the latter case~\cite[Section 10]{theorem-proving-in-lean}.
Compared to Haskell's, Lean's typeclasses have few structural restrictions: notably, classes and instances can depend on any term, instances may overlap, classes can apply to multiple types and can have functional dependencies.

In our development, we followed the common practice in \mathlib of providing structure on a type,
whenever such a structure exists canonically, through typeclasses.
The informal notion of providing a certain mathematical structure on a type should not be confused with the \lean{structure} keyword formally declaring a \emph{structure type} whose elements are tuples.
To confuse matters further, Lean implements typeclasses as structure types, where the typeclass instances are tuples of the typeclass's fields.
Typeclasses provided us a way to treat uniformly situations that are informally considered the same, as we discuss in Sections \ref{sec:field_extension} and \ref{sec:scalar_tower}.
Our reliance on typeclasses did not cause any noticeable slowness in proof checking:
there was no instance that should be found but could not due to timeouts.

A central consideration in formalizing definitions for \mathlib is choosing the appropriate amount of \emph{bundling}:
determining whether information about an object should be carried by the object itself (bundled), or passed as a separate value (unbundled)~\cite{instance-parameters-mathlib}.
For example, the \lean{is\_number\_field} typeclass of Section~\ref{sec:number fields} is considered to have unbundled inheritance from the \lean{field} class because instances of these classes are passed in separate parameters,
while it has bundled inheritance from \lean{char\_zero} and \lean{finite\_dimensional} since both are included as fields of the structure.
Similarly, the formalization of admissible absolute values discussed in Section~\ref{subsec:finiteness} features a bundled structure \lean{absolute\_value} which includes a map along with proofs stating that this map is an absolute value,
and an unbundled structure \lean{is\_admissible} which takes the absolute value map as a separate parameter.

Unbundling has an advantage in expressivity: because each property of an object is passed in a separate parameter,
modifying one hypothesis requires modifying one parameter.
In contrast, bundling hypotheses means that each subset of hypotheses requires its own structure declaration;
any results proved for a given structure have to be made available for other structures manually or through automation such as typeclass inference.
The advantage here is that bundled structures result in simpler parameter lists, since fully unbundling the \lstinline{field} class would result in each of its 38 structure fields becoming a separate parameter.

Technical considerations play another important role in choosing the level of bundling:
bundled properties are easily found by automation compared to unbundled properties which require a search of the local context,
bundled inheritance between classes can only be applied when the two classes have the same type parameters,
while long unbundled inheritance chains cause exponentially large terms, resulting in slowdowns and high memory consumption.
Although there is no general rule governing bundling, in general \mathlib prefers to bundle if possible,
unbundling only when the additional properties are all \lean{Prop}-valued and are not involved in long inheritance chains.

\section{Number fields, global fields and rings of integers} \label{sec:number fields}

We refer the reader to Section~\ref{sec math background} for the mathematical background needed in this section.

We formalized number fields as the following typeclass:
\begin{lstlisting}
class is_number_field (K : Type*) [field K] : Prop :=
[to_char_zero : char_zero K]
[to_finite_dimensional : finite_dimensional ℚ K]
\end{lstlisting}
The \emph{class} keyword declares a structure type (in other words, a type of record) and enables typeclass inference for terms of this type;
we describe the use of typeclasses in \mathlib in Section \ref{sec:typeclasses}.
Round brackets mark parameters that must explicitly be supplied by the user, such as \lean{(K : Type*)}, while
square brackets mark instance parameters inferred by the typeclass system, such as \mbox{\lean{[field K]}}.
The condition \lean{[to\_char\_zero : char\_zero K]} states that $K$ has characteristic zero, so the unique ring homomorphism $\Z \to K$ is an embedding.
This implies that there is a $\QQ$-algebra structure on $K$ (found by typeclass instance synthesis), endowing $K$ with the $\QQ$-vector space structure used in the hypothesis \mbox{\lean{[to\_finite\_dimensional : finite\_dimensional ℚ K]}}.

Similarly, we defined the class of function fields over a finite field $\Fq$ as:
\begin{lstlisting}
class function_field (Fq F : Type*) [field Fq] [field F] :
  Prop :=
[to_algebra : algebra (ratfunc Fq) F]
[to_finite_dimensional : finite_dimensional (ratfunc Fq) F]
\end{lstlisting}
The hypothesis \lean{[to\_algebra : algebra (ratfunc $\Fq$) F]} witnesses that $F$ is a field extension of the field $\Fq[q](t)$ of rational functions over $\Fq[q]$, where $\Fq[q]$ is any field (although in our applications we will insist that $\Fq[q]$ be actually finite).
Again, the condition that this extension is finite is written using the \lean{finite\_dimensional} typeclass.
We present a more detailed analysis of \lean{algebra} in Section \ref{sec:field_extension} and of fraction fields including \lean{ratfunc} in Section \ref{subsection : fields of fractions}.
For now, we point out that there are many fields $K$ that are isomorphic to the field of rational functions $\Fq[q](t)$; we provided a theorem \lean{function\_field\_iff} that shows that the choice of $K$ does not matter.
Note that there is no requirement that the field \lean{Fq} is finite, since this is not needed to state the conditions on \lean{F}.
We instead add a \lean{[fintype Fq]} hypothesis only to those results that require finiteness.

\subsection{Field extensions} \label{sec:field_extension}

The definition of \lean{is\_number\_field} illustrates our treatment of field extensions.
A field $L$ containing a subfield $K$ is said to be a field extension $L / K$.
Often we encounter towers of field extensions: we might have that $\QQ$ is contained in~$K$,~$K$ is contained in $L$, $L$ is contained in an algebraic closure $\overline{K}$ of $K$, and~$\overline{K}$ is contained in $\C$.
We might formalize this situation by viewing $\QQ$, $K$, $L$ and $\overline{K}$ as sets of complex numbers $\C$ and defining field extensions as subset relations between these subfields.
This way, no coercions need to be inserted in order to map elements of one field into a larger field.
Unfortunately, we can only avoid coercions as far as we are able to stay within one largest field.
For example, the definition of complex numbers depends on many results for rational numbers, which would need to be proved again, or transported, for the subfield of $\C$ isomorphic to $\QQ$.

Instead, we formalized results about field extensions through parametrization. The fields $K$ and $L$ can be arbitrary types
and the hypothesis ``$L$ is a field extension of~$K$'' is represented by an instance parameter \lean{[algebra K L]} denoting a $K$-algebra structure on $L$.
The \lean{algebra} structure provides us with a ring homomorphism $\lean{algebra\_map K L} : K \to L$; this map is injective because $K$ and $L$ are fields.
In other words, field extensions are given by their embeddings.

There are multiple possible $K$-algebra structures for a field $L$ and Lean does not enforce uniqueness of typeclass instances,
but the \mathlib maintainers try to ensure all instances that can be inferred are \emph{definitionally equal}.
Definitional equality is a syntactical notion of equality found in dependent type theories that reflects the possibility of computation:
for example, the term \lean{2 + 2 : ℕ} is definitionally equal to \lean{4}.
Whenever Lean can infer the definitional equality of two terms (the terms are said to \emph{unify}),
one can be substituted for the other.
Thus, ensuring definitional equality for instances means that overlapping instances will not lead to conflicts when one instance is expected and another is found.

\subsection{Scalar towers} \label{sec:scalar_tower}

The main drawback of using arbitrary embeddings to represent field extensions is that we need to prove that these maps commute.
For example, we might start with a field extension $L / \QQ$, then define a subfield $K$ of $L$,
resulting in a tower of extensions $L / K / \QQ$.
In such a tower, the map $\QQ \to L$ should be equal to the composition $\QQ \to K$ followed by $K \to L$.
Such an equality cannot always be achieved by defining the map $\QQ \to L$ to be this composition: in the example, the definition of the map $\QQ \to K$ depends on the map $\QQ \to L$.

The solution in \mathlib is to parametrize over all three maps, as long as there is also a proof of coherence:
a hypothesis of the form ``$L / K / F$ is a tower of field extensions'' is translated into three instance parameters \lean{[algebra F K]}, \lean{[algebra K L]} and \lean{[algebra F L]},
along with a parameter \lean{[is\_scalar\_tower F K L]} expressing that the maps commute.

The \lean{is\_scalar\_tower} typeclass derives its name from its applicability to any three types among which scalar multiplication operations exist:
\begin{lstlisting}
class is_scalar_tower (M N α : Type*)
  [has_scalar M N] [has_scalar N α] [has_scalar M α] : Prop :=
(smul_assoc : ∀(x : M) (y : N) (z : α), (x • y) • z = x • (y • z))
\end{lstlisting}
For example, if $R$ is a ring, $A$ is an $R$-algebra and $M$ an $A$-module, we can state that $M$ is also an $R$-module by adding a \lean{[is\_scalar\_tower R A M]} parameter.
Since \lean{x~$\cdot$~y} for an $R$-algebra $A$ is defined as \lean{algebra\_map R A x * y}, applying \lean{smul\_assoc} for each $x : K$ with $y = (1 : L)$ and $z = (1 : F)$ shows that the \lean{algebra\_map}\kern-.3em s indeed commute in a tower of field extensions $L / K / F$.

Common \lean{is\_scalar\_tower} instances are declared in \mathlib,
such as for the maps $R \to S \to B$ when $S$ is a $R$-subalgebra of $A$ and $B$ is an $A$-algebra such that \lean{is\_scalar\_tower R A B};
this also implies that the maps $R \to S \to A$ form a tower.
The effect is that almost all coherence proof obligations are automated through typeclass instance synthesis.
Only when defining a new algebra structure were we required to supply the \lean{is\_scalar\_tower} instances ourselves.

\subsection{Rings of integers} \label{sec:ring-of-integers}

When $K$ is a number field (defined as a field satisfying \lean{is\_number\_field}), the ring $\OK$ of integers in $K$ is defined as the integral closure of $\Z$ in $K$.
This is the subring containing those $x : K$ that are a root of a monic polynomial with coefficients in $\Z$:
\begin{lstlisting}
def number_field.ring_of_integers (K : Type*) [field K]
  [is_number_field K] : subalgebra ℤ K :=
integral_closure ℤ K
\end{lstlisting}
where \lean{integral\_closure} was already defined in \mathlib.
When $K$ is a function field over the finite field $\Fq$, we defined $\OK$ analogously as \lean{integral\_closure ($\Fq$[X]) K}.

Since the integers $\Z$ are integrally closed in $\QQ$, this construction of the ring of integers of the number field $\QQ$ is isomorphic, but not definitionally equal, to $\Z$.
To avoid dealing with these isomorphisms, and also to treat the two definitions of rings of integers on an equal footing,
we introduced a typeclass \lean{is\_integral\_closure A R B} stating that $A$ is the integral closure of $R$ in~$B$,
and worked with a generic \lean{is\_integral\_closure} instance instead of the specific \lean{ring\_of\_integers} construction when possible.

\subsection{Subobjects} \label{sec:subobjects}

The ring of integers is one example of a subobject, such as a subfield, subring or subalgebra, defined through a characteristic predicate.
In \mathlib, subobjects are ``bundled'', in the form of a \lean{structure} comprising the carrier set and proofs showing the carrier set is closed under the relevant operations.
Bundled subobjects provide similar benefits to those of bundled morphisms; the choice for the latter is explained in the \mathlib overview paper~\cite{mathlib}.
Where the \lean{algebra} and \lean{is\_scalar\_tower} typeclasses provide an interface generalizing over multiple equivalent definitions, subobjects provide a specific implementation of that interface in the form of a subtype.

Two new subobjects that we defined in our development were \lean{subfield} as well as \lean{intermediate\_field}. We defined a subfield of a field $K$ as a subset of~$K$ that contains $0$ and $1$ and is closed under addition, negation, multiplication and taking inverses.
If $L$ is a field extension of~$K$, we defined an intermediate field as a subfield of $L$ that is also a $K$-subalgebra: in other words, a subfield that contains the image of $\lean{algebra\_map K L}$.
Other examples of subobjects available in \mathlib are submonoids, subgroups and submodules (with ideals as a special case of submodules);
all of these are provided with an instance of the \lean{set\_like} typeclass that supplies notation such as a membership relation ``$x \in S$''.

The new definitions found immediate use:
soon after we contributed our definition of \lean{intermediate\_field} to \mathlib,
the Berkeley Galois theory group used it in a formalization of the primitive element theorem.
Soon after the primitive element theorem was merged into \mathlib,
we used it in our development of the trace form.
This anecdote illustrates the decentralized development style of \mathlib,
with different groups and people building on each other's results in a collaborative process.

Through the \lean{set\_like} typeclass, subobjects can be coerced to types, by sending a subobject $S$ to the subtype of all elements of $S$.
By putting typeclass instances on this subtype,
we could reason about inductively defined rings such as $\Z$ and subrings such as \lean{integral\_closure $\Z$ K} uniformly.
If $S : \lean{subfield}\ K$, there is a ring embedding, the map that sends $x : S$ to $K$ by ``forgetting'' that $x \in S$,
and we registered this map as an \lean{algebra S K} instance, also allowing us to treat field extensions of the form $\QQ \to \C$ and subfields uniformly.
Similarly, for $F : \lean{intermediate\_field K L}$, we defined the corresponding \lean{algebra K F}, \lean{algebra F L} and \lean{is\_scalar\_tower K F L} instances.

\subsection{Fields of fractions}\label{subsection : fields of fractions}
The fraction field $\Frac R$ of an integral domain $R$ can be defined explicitly as a quotient type as follows:
starting from the type of pairs $(a,b)$ with $a,b \in R$ such that $b\neq 0$,
one quotients by the equivalence relation generated by $(a,b) \sim (a \alpha, b \alpha)$ for all $\alpha \ne 0 : R$, writing the equivalence class of $(a,b)$ as $\frac{a}{b}$.
It can easily be proved that the ring structure on $R$ extends uniquely to a field structure on $\Frac R$;
in \mathlib this construction is called \mbox{\lean{fraction\_ring R}},
and is used to define the field of rational functions $K(X) = \lean{ratfunc K}$.
When $R=\Z$, this yields the traditional description of $\QQ$ as the set of equivalence classes of fractions, where $\frac{2}{3}=\frac{-4}{-6}$, etc.

The drawback of this construction is that there are many other fields that can serve as the field of fractions for the same ring.
Consider the field $\{z \in \C : \Re z \in \QQ, \Im z\in\QQ\}$, which is isomorphic to $\Frac (\Z[i])$ but not definitionally equal to it.
Indeed, the \mathlib definition of the rational numbers $\QQ$ is a product type, not a quotient type, so we would not be able to treat $\QQ$ as the field of fractions of $\Z$ in this setup.
Any properties proven for $\QQ$ would have to be repeated for $\Frac(\Z)$,
using transfer lemmas stating these properties are preserved by the isomorphism between $\QQ$ and $\Frac(\Z)$.

The strategy used in \mathlib is to rather allow for many different \emph{fraction fields} of our given integral domain $R$ --- as fields $K$ with a suitable \lean{[algebra R K]} instance, where the map \lean{algebra\_map R K} witnesses that all elements~$K$ are ``fractions'' of elements of $R$ --- and to parametrize every result over the choice of $K$.
The conditions on the $R$-algebra structure on $K$ are encoded as a typeclass \lean{is\_fraction\_ring R K}.
In the definition used by \mathlib, a fraction ring is a special case of a \emph{ring localization},
which is defined for any commutative ring $R$.
Different localizations restrict the denominators to different multiplicative submonoids of $R\setminus\{0\}$.

The conditions on \lean{algebra\_map R K} imply that $K$ is the smallest field, up to isomorphism, containing $R$, expressed by the following unique mapping property.
If $g \colon R \to A$ is an injective map to a ring $A$ such that $g(x)$ has a multiplicative inverse for all $x \ne 0 : R$, then
it can be extended uniquely to a map $K \to A$ compatible with \lean{algebra\_map R K} and $g$.
In particular, given \lean{is\_fraction\_ring R K₁} and \lean{is\_fraction\_ring R K₂}, we can derive an isomorphism $K_1 \simeq K_2$.
The construction of $\Frac R$ then results in \emph{a} field of fractions (with an instance \lean{is\_fraction\_ring R (fraction\_ring R)}) rather than \emph{the} field of fractions.

The above description of fraction fields is the third such formalization in \mathlib. The first version
consisted of a quotient type \lean{quotient\_ring R},
constructed similarly to the current definition of \lean{fraction\_ring R}.
Due to the aforementioned drawback --- namely, that this provided no easy way to view~$\QQ$ as the field of fractions of $\Z$, for instance ---
this was refactored to use a characteristic predicate instead.

The second version
defined $K$ to be the field of fractions of $R$ if there existed an injective \emph{fraction map} $f : R \to K$, which is a ring homomorphism witnessing that all elements of $K$ are ``fractions'' of elements of $R$;
the map and its properties were bundled as a type \lean{fraction\_map R K}.
Results on fraction fields were parametrized over a choice of fraction map $f$.
This made it possible to view $\QQ$ as the fraction field of $\Z$, by providing a suitable map called \lean{int.fraction\_map : fraction\_map ℤ ℚ}.
This came at a price:
informally, at any given stage of one's reasoning, the field $K$ is fixed and the map $f\colon R\to K$ is applied implicitly, just viewing every $x:R$ as $x:K$.
It is now impossible to view $R \leq K$ as an inclusion of $R$-subalgebras,
because the map $f$ is needed explicitly to give the $R$-algebra structure on $K$.
As a workaround, \mathlib used a type synonym \lean{codomain f := K} and instantiated the $R$-algebra structure given by $f$ on this synonym.
Again we encountered a distinction between $\QQ$ ``itself'' and $\Frac(\Z) = \lean{codomain int.fraction\_map}$,
still requiring the transfer of results such as typeclass instances.

The most recent version is the one described above.
Inspired by our success in using the \lean{algebra} typeclass to denote inclusions of rings,
we unbundled the explicit \lean{(f : fraction\_map R K)} parameters into
an instance parameter \lean{[algebra R K]} that specifies the map,
and an instance parameter \lean{[is\_fraction\_ring R K]} that specifies the conditions satisfied by the map.
Separating out these parameters finally allowed us to painlessly view $\QQ$ as the fraction ring of $\Z$ while preserving the original $\Z$-algebra structure on $\QQ$.

\subsection{Representing monogenic field extensions} \label{sec:monogenic-field-extension}

In Section~\ref{sec math background} we have informally said that every number field $K$ can be written as $K=\QQ(\alpha)$ for a root $\alpha$ of an irreducible polynomial $P\in\QQ[X]$. This can be made precise in several ways. For instance, one can consider a large field~$L$ (of characteristic~$0$) where $P$ splits completely, then choose a root $\alpha\in L$ and let $K = \QQ(\alpha)$ be the smallest subfield of $L$ containing $\alpha$. Or, one can consider the quotient ring $\QQ[X]/P$ and observe that this is a field where the class $X\pmod{P}$ is a root of $P$. The assignment $\alpha\mapsto X\pmod{P}$ yields an isomorphism of the two fields, but any other choice of a root $\alpha'\in L$ leads to another isomorphism $\QQ(\alpha')\cong \QQ[X]/P$. Although mathematically we often tacitly identify these constructions, there is no canonical representation of the \emph{monogenic} extensions of $\QQ$, those which can be obtained by adjoining a single root of one polynomial.

The same continues to hold if we replace the base field $\QQ$ with another field~$F$, thus considering extensions of the form $F(\alpha)$, now requiring that $\alpha$ be a root of some $P\in F[X]$. Various constructions of $F(\alpha)$ have already been formalized in \mathlib. The ability to switch between these representations is important: sometimes $K$ and $F$ are fixed and we want an arbitrary $\alpha$; sometimes $\alpha$ is fixed and we want an arbitrary type representing $F(\alpha)$.

To find a uniform way to reason about all these definitions,
we chose to formalize the notion of \emph{power basis} to represent monogenic field extensions: this is a basis of the form $1, \alpha, \alpha^2, \dots, \alpha^{n-1} : K$ (viewing $K$ as a $F$-vector space).
We defined a structure type bundling the information of a power basis.
Omitting some generalizations not needed in this paper, the definition reads:
\begin{lstlisting}
structure power_basis (F K : Type*) [field F] [field K]
  [algebra F K] :=
(gen : K) (dim : ℕ) (basis : basis (fin dim) F K)
(basis_eq_pow : ∀ i, basis i = gen ^ (i : ℕ))
\end{lstlisting}
We formalized that the previously defined notions of monogenic field extensions are equivalent to the existence of a power basis.

With the \lean{power\_basis} structure, we gained the ability to parametrize our results,
being able to choose the $F$ and $K$ in a monogenic field extension $K / F$,
or being able to choose the $\alpha$ generating $F(\alpha)$ (by setting the \lean{gen} field to $\alpha$).
To specialize a result from an arbitrary $K$ with a power basis over $F$ to a specific construction of $K = F(\alpha)$,
one can apply the result to the power basis \lean{pb} generated by $\alpha$ and rewrite $\lean{power\_basis.gen pb} = \alpha$.

\section{Dedekind domains} \label{sec:Dedekind-domain}
The right setting to study algebraic properties of number fields are \emph{Dedekind domains}.
We formalized fundamental results on Dedekind domains, including the equivalence of two definitions of Dedekind domains.

\subsection{Definitions}\label{subsec:definitions_DD}
There are various equivalent conditions, used at various times, for an integral domain $D$ to be a Dedekind domain.
The following three have been formalized in \mathlib:
\begin{itemize}
\item \lean{is\_dedekind\_domain D}: $D$ is a Noetherian integral domain, integrally closed in its fraction field and has Krull dimension at most $1$;
\item \lean{is\_dedekind\_domain\_inv D}: $D$ is an integral domain and nonzero fractional ideals of $D$ have a multiplicative inverse (we discuss the notion and formalization of fractional ideals in Section~\ref{subsection:frac_ideals});
\item \lean{is\_dedekind\_domain\_dvr D}: $D$ is a Noetherian integral domain and the localization of $D$ at each nonzero prime ideal is a discrete valuation ring.
\end{itemize}
Note that fields are Dedekind domains according to these conventions.

The mathlib community chose \lean{is\_dedekind\_domain} as the main definition,
since this condition is usually the one checked in practice~\cite{Neukirch}.
The other two equivalent definitions were added to \mathlib, but before formalizing the proof that they are indeed equivalent.
Having multiple definitions allowed us to do our work in parallel without depending on unformalized results.
For example,
the proof of unique ideal factorization in a Dedekind domain initially assumed \lean{is\_dedekind\_domain\_inv D},
and the proof that the ring of integers $\OK$ is a Dedekind domain concluded \lean{is\_dedekind\_domain (ring\_of\_integers K)}.
After the equivalence between \lean{is\_dedekind\_domain D} and \lean{is\_dedekind\_domain\_inv D} was formalized,
we could easily replace usages of \lean{is\_dedekind\_domain\_inv} with \lean{is\_dedekind\_domain}.

The conditions \lean{is\_dedekind\_domain} and \lean{is\_dedekind\_domain\_inv} require a fraction field $K$,
although the truth value of the predicates does not depend on the choice of $K$.
For ease of use, we let the type of \lean{is\_dedekind\_domain} depend only on the domain $D$
by instantiating $K$ in the definition as \lean{fraction\-\_ring D}. From now on, we fix a fraction field $K$ of $D$.
\begin{lstlisting}
class is_dedekind_domain (D : Type*)
  [comm_ring D] [is_domain D] :=
(is_noetherian_ring : is_noetherian_ring D)
(dimension_le_one : dimension_le_one D)
(is_integrally_closed : is_integrally_closed D)
\end{lstlisting}

Applications of \lean{is\_dedekind\_domain} can choose a specific fraction field through the following lemma exposing the alternate definition:
\begin{lstlisting}
lemma is_dedekind_domain_iff [is_fraction_ring D K] :
  is_dedekind_domain D ↔
    is_noetherian_ring D ∧ dimension_le_one D ∧
    ∀ {x : K}, is_integral D x →
      ∃ (y : D), algebra_map D K y = x
\end{lstlisting}

We marked \lean{is\_dedekind\_domain} as a typeclass by using the keyword \lean{class} rather than \lean{structure},
allowing the typeclass system to automatically infer the Dedekind domain structure when an appropriate instance is declared, such as for PIDs or for rings of integers.

\subsection{Fractional ideals}\label{subsection:frac_ideals}
The notion which is pivotal to the definition of the ideal class group of a Dedekind domain is that of \emph{fractional ideals}:
given any integral domain $R$ with a field of fractions $F$,
we define \lean{is\_fractional} as a predicate on $R$-submodules $J$ of $F$, informally as ``there is an $x : R$ with $x J \subseteq R$''.
For a Dedekind domain, nonzero fractional ideals form a group under multiplication.
As seen in Section~\ref{subsection : fields of fractions}, this notion depends on the field $F$ as well as on the embedding \lean{$f$ := algebra\_map R F}.
A more precise way of stating the above condition is then
$f(x)J\subseteq f(R)$.
We formalized the definition of fractional ideals of $R$ contained in $F$ as a type \lean{fractional\_ideal R F}, whose elements consist of an $R$-submodule of $F$ along with a proof of \lean{is\_fractional}.
The structure of fractional ideals does not depend on the choice of a fraction field,
which we formalized as an isomorphism \lean{fractional\_ideal.canonical\_equiv} between two types of fractional ideals on $R$, corresponding to different fields of fractions.

We defined the addition, multiplication and intersection operations on fractional ideals,
by showing that the corresponding operations on submodules map fractional ideals to fractional ideals.
We also formalized that these operations give a commutative semiring structure on the type of fractional ideals.
For example, multiplication of fractional ideals is defined as
\begin{lstlisting}
lemma is_fractional.mul (I J : submodule R F)
  is_fractional R I → is_fractional R J →
  is_fractional R (I * J) := _ -- proof omitted

instance : has_mul (fractional_ideal R F) :=
⟨λ I J, ⟨I * J : submodule R F,
 is_fractional.mul I.is_fractional J.is_fractional⟩⟩
\end{lstlisting}

Defining the quotient of two fractional ideals requires slightly more work. Consider any $R$-algebra $A$ and an injection $R\hookrightarrow A$. Given ideals $I,J\le R$, the submodule $I / J\le A$
is defined by the property
\pagebreak[3] 
\begin{lstlisting}
lemma submodule.mem_div_iff_forall_mul_mem {x : A}
  {I J : submodule R A} :
  x ∈ I / J ↔ ∀ y ∈ J, x * y ∈ I
\end{lstlisting}
Beware that the notation $1/I$ might be misleading here: indeed, for general integral domains, the equality $I\ast 1/I=1$ might not hold. As an example, one can consider the ideal $(X,Y)$ in $\C[X,Y]$, which is not a Dedekind domain: by definition, $(X, Y)^{-1}$ consists of the elements $a=\frac{p}{q} \colon \Frac \bigl(\C[X,Y]\bigr)$ with the property that $a \ast b \in \C[X,Y]$ for all $b \in (X,Y)$. This last condition is equivalent to requiring that both $a \ast X$ and $a \ast Y$ are in $\C[X,Y]$ and thus the denominator $q$ of $a$ must be divisible both by $X$ and by $Y$, so actually $q\in\C^\times$. It follows that $(X,Y)^{-1}=\C[X,Y]$, and in particular $(X,Y)\ast (X,Y)^{-1}=(X,Y)\subsetneq 1=\C[X,Y]$.

On the other hand, we formalized that the equality $I\ast 1/I=1$ holds for Dedekind domains (Section~\ref{sec:equivalence}) as the following lemma:
\begin{lstlisting}
theorem fractional_ideal.mul_inv_cancel [is_dedekind_domain D]
  {I : fractional_ideal D F} (hne : I ≠ 0) : I * (1 / I) = 1
\end{lstlisting}
This justifies the notation $I^{-1}=1/I$. In fact, we define this notation even for the ideal $0$, by declaring that $0^{-1}=0$.
This fits the pattern of the typeclass \lean{group\_with\_zero} in \mathlib, consisting of groups endowed with an extra element \lean{0} whose inverse is again \lean{0}.

Moreover, \mathlib used to define \lean{$a / b := a * b^{-1}$}, but our definition of $I^{-1} = 1 / I$ would cause circularity. This led us to a major refactor of this core definition. In particular, we had to weaken the definitional equality to a proposition; this involved many small changes throughout \mathlib
\footnote{The pull requests are available as \url{https://github.com/leanprover-community/mathlib/pull/5302} and \url{https://github.com/leanprover-community/mathlib/pull/5303}.}.

\subsection{Equivalence of the definitions} \label{sec:equivalence}
We now describe how we proved and formalized that the two definitions \lean{is\_dedekind\_domain} and \lean{is\_dedekind\_domain\_inv} of being a Dedekind domain are equivalent. Let $D$ be a Dedekind domain, and let $f\colon D\to K$ a fraction map to a field of fractions $K$ of $D$.

To show that \lean{is\_dedekind\_domain\_inv} implies \lean{is\_dedekind\_domain}, we follow the proof given by Fr\"ohlich in~\cite[Chapter 1, Section~2, Proposition~1.2.1]{frohlich}. A constant challenge that was faced while coding this proof was already mentioned in Section \ref{subsection : fields of fractions}, namely the fact that elements of the domain must be traced along the inclusion into the chosen field of fractions.
The proofs for being integrally closed and of dimension being less than or equal to $1$ are fairly straightforward.

Formalizing the Noetherian condition was the most challenging. Fr\"ohlich considers elements $a_1, \dots, a_n \in I$ and $b_1, \dots, b_n \in I^{-1}$ for any nonempty fractional ideal $I$,
satisfying $ \sum_i a_i b_i = 1 $. Observe now that, in \mathlib, the definition of the product $A\ast B$ of two fractional ideals $A,B$ is a special case of the product of two submodules, and therefore it is defined as
\begin{lstlisting}
submodule.has_mul = {mul := $\lambda$ (A B : submodule D K),
 $\sqcup$ (a : A), submodule.map ((algebra.lmul D K) a.val) B}
\end{lstlisting}
Unraveling this definition, we see that it defines $A\ast B$ as the smallest (\emph{i.~e.}, the infimum with respect to set-theoretic inclusion as order relation) submodule containing all submodules $a\cdot B$ for $a\in A$, where $a \cdot B$ is the range of the function \lean{$\lambda$ b : B, $a\ast b$}. However, it is quite challenging to formalize that an element of $A\ast B$ must be a \emph{finite} sum $\sum_{i} a_i*b_i$, for $a_i \in A$ and $b_i \in B$. Instead, we show that, for every element $x\in A\ast B$, there are finite sets $T\subseteq A$, $T'\subseteq B$ such that \lean{x $\in$ span (T * T')}, formalized as
\lean{submodule.mem\_span\_mul\_finite\-\_of\-\_mem\-\_mul}.
Now considering a nonzero integral ideal $I$ of the ring $D$, by definition of invertibility we can write \lean{1 $\in$ (1 $:$ fractional\_ideal D K) = I * I\inv}. Hence, we obtain 
finite sets $T \subset I$ and $T' \subset I\inv$ such that $1$ is contained in the $D$-span of $T*T'$. We used the \lean{norm\_cast} tactic~\cite{norm_cast} to resolve most coercions but this tactic did not solve coercions coming from the inclusion \lean{algebra\_map D K}. With coercions, the actual statement of the latter expression in Lean is \lean{↑T' ⊆ ↑↑(↑I)\inv}, which reads
\begin{lstlisting}
(T' : set K) ⊆ (((I : fractional_ideal D K)⁻¹ : submodule D K) : set K)
\end{lstlisting}
From the existence of $T$ and $T'$ we concluded that~$I$ is indeed finitely generated, thus finishing the proof.

The theorem \lean{fractional\_ideal.mul\_inv\_cancel} proves the converse, namely that \lean{is\_dedekind\_domain} implies \lean{is\_dedekind\_domain\_inv}. The classical proof consists of three steps: first, every maximal ideal $M\subseteq D$, seen as a fractional ideal, is invertible;
second, every nonzero ideal is invertible, using that it is contained in a maximal ideal;
third, the fact that every fractional ideal $J$ satisfies $xJ\leq I$ for a suitable element $x\in D$ and a suitable ideal $I\subseteq D$ implies that every fractional ideal is invertible, concluding the proof that nonzero fractional ideals form a group.
The third step was easy, building upon the material developed for the general theory of \lean{fractional\_ideal}. Concerning the first two, we found that passing from the case where $M$ is maximal to the general case required more code than directly showing invertibility of arbitrary nonzero ideals. The formal statement reads
\begin{lstlisting}
lemma coe_ideal_mul_inv [is_dedekind_domain D] 
  (I : ideal D) (hI0 : I ≠ 0) :
  (↑I * (↑I)⁻¹ : fractional_ideal D K) = 1
\end{lstlisting}
from where it becomes apparent that we had to repeatedly distinguish between \lean{I $:$ ideal D}, and its coercion \lean{↑I $:$ fractional\_ideal D K} although these objects, from a mathematical point of view, are identical.

The formal proof of this result relies on the lemma \lean{exists\_not\_mem\_one\-\_of\-\_ne\-\_bot}, which says that for every non-trivial ideal $0\subsetneq I\subsetneq D$, there exists an element in the field $K$ which is not integral (so, not in \lean{1 : fractional\_ideal D K}) but lies in $I\inv$. The proof begins by invoking that every nonzero ideal in the Noetherian ring $D$ contains a product of nonzero prime ideals. This result was not previously available in \mathlib.
The dimension condition shows its full force when applying this lemma:
each prime ideal in the product $I\ast I^{-1}$, being nonzero, will be maximal because the Krull dimension of $D$ is at most $1$; from this, \lean{exists\_not\_mem\_one\_of\_ne\_bot} follows easily. Having the above lemma at our disposal,
we were able to prove that every ideal $I\ne 0$ is invertible by arguing by contradiction: if $I\ast I\inv \ne D$, we can find an element $x\in K\setminus D$ which is in $(I\ast I\inv)\inv$ thanks to \lean{exists\_not\_mem\_one\_of\_ne\_bot}; some easy algebraic manipulation then implies that $x$ is actually integral over $D$. Since $D$ is integrally closed, $x \in D$, contradicting the construction of $x$. Combining these results gives the equivalence between the two conditions for being a Dedekind domain.

\subsection{Unique ideal factorization}\label{subsec:unique_ideal_factorization}

As briefly indicated before, we also formalized a proof that in a Dedekind domain every nonzero ideal can be expressed as a product of prime ideals in a unique way up to the order of the factors.
In fact, for an integral domain, every nonzero ideal is a product of prime ideals if and only if all nonzero fractional ideals are invertible; the uniqueness follows separately. We have formalized one direction of this equivalence, a proof of the converse can be found in~\cite[Chapter 5, \textsection 6, Theorem 10]{Zariski-Samuel}.

We formalized the unique ideal factorization property of a Dedekind domain $D$ by instantiating a unique factorization monoid structure on its ideals.

\begin{lstlisting}
instance ideal.unique_factorization_monoid :
  unique_factorization_monoid (ideal D)
\end{lstlisting}

In \mathlib, unique factorization domains are actually a special case of unique factorization monoids (UFMs).
A commutative monoid $R$ with an absorbing element $0$ and injective multiplication is defined to be a UFM,
if the relation ``$x$ properly divides $y$'' is well-founded (implying that every element can be factored as a product of irreducibles) and
an element of $R$ is prime if and only if it is irreducible (implying uniqueness of the factorization).
Examples in \mathlib of UFMs are the unique factorization domains $\N$ and $\Z$ as well as, for any UFM \lean{α}, the quotient of \lean{α} by the subgroup of invertible elements \lean{associates α}.
With much of the necessary definitions and properties already formalized, the formalization of this unique factorization result has been done in well under 100 lines of Lean code. One of the main mathematical ingredients (interesting in its own right) is that for ideals in a Dedekind domain, to divide is to contain:

\begin{lstlisting}
lemma ideal.dvd_iff_le {I J : ideal D} : (I ∣ J) ↔ J ≤ I
\end{lstlisting}

Similarly, to strictly contain is to properly divide, so the well-foundedness condition of UFMs is exactly the property that a Dedekind domain is Noetherian. In order to show that all irreducible elements of the monoid of nonzero prime ideals in $D$ are prime elements, we formalized that irreducible ideals in a Dedekind domain are maximal and therefore prime (note that prime ideals of a Dedekind domain $D$ coincide with prime elements of the monoid of its nonzero ideals); the converse holds in every monoid.

We note that the unique factorization result, or actually an easy corollary thereof, is an important ingredient in our finiteness proof for the class group of rings of integers, as we will elaborate on in Section~\ref{subsec:finiteness}.

\section{Principal ideal domains are Dedekind}

As an example of our definitions, we discuss in some detail our formalization of the fact that a principal ideal domain is a Dedekind domain.
In the same way that unique factorization domains are generalized in \mathlib to unique factorization monoids,
there is no explicit definition of PIDs in \mathlib. Rather, it is split up into multiple hypotheses.
One uses \lean{[comm\_ring R] [is\_domain R] [is\_principal\_ideal\_ring R]} to denote a PID $R$,
where \lean{is\_domain} is a typeclass asserting that the ring is nontrivial and there are no zero divisors,
and \lean{is\_principal\_\-ideal\_\-ring} is a typeclass defined for all commutative rings:
\begin{lstlisting}
class is_principal_ideal_ring (R : Type*) [comm_ring R] :=
(principal : ∀ (I : ideal R), is_principal I)
\end{lstlisting}

Our proof that the hypotheses \lean{[comm\_ring R] [is\_domain R] [is\_principal\_\-ideal\_\-ring R]} imply \lean{is\_dedekind\_domain R} was relatively short:
\begin{lstlisting}
instance principal_ideal_ring.to_dedekind_domain (R : Type*)
  [comm_ring R] [is_domain R] [is_principal_ideal_ring R] :
  is_dedekind_domain R :=
⟨principal_ideal_ring.is_noetherian_ring,
 dimension_le_one.principal_ideal_ring R,
 unique_factorization_monoid.is_integrally_closed⟩
\end{lstlisting}

The Noetherian property of a Dedekind domain followed easily by the previously defined lemma \lean{principal\_ideal\_ring.is\_noetherian\_ring}, since, by definition, each ideal in a principal ideal ring is finitely generated (by a single element).

We proved the lemma \lean{dimension\_le\_one.principal\_ideal\_ring}, which is an instantiation of the existing result \lean{is\_prime.to\_maximal\_ideal}, showing that a nonzero prime ideal in a PID is maximal.
The latter lemma uses the characterization that $I$ is a maximal ideal if and only if any strictly larger ideal $J\supsetneq I$ is the full ring $R$.
If $I$ is a nonzero prime ideal and $J \supsetneq I$ in the PID $R$, we see that a generator $j$ of $J$ is a divisor of any generator $i$ of $I$. Since $I$ is prime, this implies that either $j \in I$, contradicting the assumption that $J \supsetneq I$, or $i = 0$, contradicting that $I$ is nonzero, or finally that $j$ is a unit, implying $J = R$ as desired.

The final condition of a PID being integrally closed was the most challenging.
We used the previously defined instance \lean{principal\_ideal\_ring.to\_\-unique\-\_factorization\_monoid} to deduce that a PID is a unique factorization monoid (UFM),
to instantiate our proof that every UFD is integrally closed.
In a PID, the Noetherian property implies that the division relation is well-founded,
and \lean{principal\_ideal\_ring.irreducible\_iff\_prime} shows that irreducible elements and prime elements coincide.
To prove that an irreducible element $p$ is prime, the proof uses that prime elements generate prime ideals and irreducible elements of a PID generate maximal ideals. Since all maximal ideals are prime ideals, the ideal generated by $p$ is maximal, hence prime, thus $p$ is prime.
We proved the lemma \lean{irreducible\_of\_prime}, which shows the converse holds in any commutative monoid with zero.

To show that a UFM is integrally closed, we first formalized the Rational Root Theorem, named \lean{denom\_dvd\_of\_is\_root},
which states that for a polynomial $p : R[X]$ and an element of the fraction field $x : \Frac R$ such that $p(x) = 0$, the denominator of $x$ divides the leading coefficient of $p$.
If $x$ is integral with minimal polynomial $p$, the leading coefficient is $1$, therefore the denominator is a unit and $x$ is an element of $R$.
This gave us the required lemma \lean{unique\_factorization\_monoid.integrally\_closed}, which states that the integral closure of $R$ in its fraction field is $R$ itself.

\section{Rings of integers are Dedekind domains} \label{sec:integral-closure}

An important classical result in algebraic number theory is that the ring of integers of a number field $K$, defined as the integral closure of $\Z$ in $K$, is a Dedekind domain. We formalized a stronger result: given a Dedekind domain~$D$ and a field of fractions $F$, if $K$ is a finite separable extension of~$F$, then the integral closure of $D$ in $K$ is a Dedekind domain with fraction field $K$.
Our approach was adapted from Neukirch \cite[Theorem~3.1]{Neukirch}.
Throughout this section, let $D$ be a Dedekind domain with a field of fractions $F$, $K$ a finite, separable field extension of~$F$ and let $S$ denote the integral closure of $D$ in $K$.

The first step was to show that $K$ is a field of fractions for the integral closure, namely, that there is an instance \lean{is\_fraction\_ring\_of\_finite\_extension D F K : is\_fraction\_ring S K}.
The main content of \lean{is\_fraction\_ring\_\-of\-\_finite\-\_extension} consisted of showing that all elements $x : K$ can be written as $y / z$ for elements $y \in S$, $z \in D \subseteq S$;
the standard proof of this fact (see~\cite[Theorem~15.29]{Dummit-and-Foote}) formalized readily.

We could then show that the integral closure of $D$ in $K$ is a Dedekind domain,
by proving it is integrally closed in $K$, has Krull dimension at most $1$ and is Noetherian.
The fact that the integral closure is integrally closed was immediate.

To show the Krull dimension is at most $1$, we needed to develop basic going-up theory for ideals.
In particular, we showed that an ideal $I$ in an integral extension is maximal if it lies over a maximal ideal,
and used a result already available in \mathlib that a prime ideal $I$ in a ring extension lies over a prime ideal.
\begin{lstlisting}
lemma is_maximal_of_is_integral_of_is_maximal_comap
  [algebra R S] (hRS : algebra.is_integral R S)
  (I : ideal S) [is_prime I]
  (hI : is_maximal (comap (algebra_map R S) I)) : is_maximal I
theorem is_prime.comap (f : R →+* S) (I : ideal S)
  [hI : is_prime I] : is_prime (comap f I)
\end{lstlisting}

The final condition, that the integral closure $S$ of $D$ in $L$ is a Noetherian ring, required the most work.
We started by following the first half of Dummit and Foote~\cite[Theorem~15.29]{Dummit-and-Foote},
so that it sufficed to find a nondegenerate bilinear form $B$ such that all integral $x, y : K$ satisfy $B(x, y) \in \lean{integral\_closure}\ D\ K$.
We then formalized the results in Neukirch \cite[Sections~2.5--2.8]{Neukirch} to show that the \emph{trace form} is a bilinear form satisfying these requirements.

\subsection{The trace form}\label{sec:trace-form}
In the notation from the previous section, consider the bilinear map \lean{lmul := $\lambda$ x y $:$ K, x~*~y}.
The trace of the linear map \lean{lmul x} is called the \emph{algebra trace} $\Tr_{K / F}(x)$ of $x$.
We defined the algebra trace
as a linear map, in this case from $K$ to $F$:
\begin{lstlisting}
noncomputable def trace : K →ₗ[F] F :=
linear_map.comp (linear_map.trace F K)
  (to_linear_map (lmul F K))
\end{lstlisting}
This definition was marked noncomputable since \lean{linear\_map.trace} makes a case distinction on the existence of a finite basis,
choosing an arbitrary finite basis if one exists (since the value of \lean{linear\_map.trace} does not depend on this choice)
and returning $0$ otherwise.
This latter case did not occur in our development.

We defined the \emph{trace form} to be an $F$-bilinear form on $K$, mapping $x, y : K$ to $\Tr_{K/F}(xy)$.
\begin{lstlisting}
noncomputable def trace_form : bilin_form F K :=
{ bilin := λ x y, trace F K (x * y), .. /- proofs omitted -/ }
\end{lstlisting}

In the following, let $L / K / F$ be a tower of finite extensions of fields, namely we assume \lean{[algebra F K] [algebra K L] [algebra F L] [is\_scalar\_tower F K L]}, as described in Section~\ref{sec:scalar_tower}.

The value of the trace depends on the choice of $F$ and $K$; we formalized this as lemmas \lean{trace\_algebra\_map x $:$ trace F K (algebra\_map F K x) = finrank F K * x} as well as \lean{trace\_trace x $:$ trace F K (trace (K L x)) = trace F L x};
here \lean{finrank F K} is the degree of the field extension $K / F$.
These results followed by direct computation.

To compute $\Tr_{K/F}(x)$, it therefore suffices to consider the trace of $x$ in the smallest field containing $x$ and $F$, which is the monogenic extension $F(x)$ discussed in Section \ref{sec:monogenic-field-extension}.
There is a nice formula for the trace in $F(x)$, although the terms in this formula are elements in a larger field $L$
(such as the \emph{splitting field} of \lean{minpoly F x}, the minimal polynomial of $x$ over $F$).
In formalizing this formula, we first mapped the trace to $L$ using the embedding $\lean{algebra\_map F L}$,
which gave the following statement:
\begin{lstlisting}
lemma power_basis.trace_gen_eq_sum_roots (pb : power_basis F K)
  (h : polynomial.splits (algebra_map F L) (minpoly F pb.gen)) :
  algebra_map F L (trace F K pb.gen) =
    sum (roots (map (algebra_map F L) (minpoly F pb.gen)))
\end{lstlisting}
We formulated the lemma in terms of the power basis, since we needed to use it for $F(x)$ here
and for an arbitrary finite separable extension $L / K$ later in the proof.

The elements of \lean{roots (map (algebra\_map F L) (minpoly F pb.gen))} are called \emph{conjugates} of $x$ in $L$.
Each conjugate of $x$ is integral since it is a root of the same monic polynomial,
and integer multiples and sums of integral elements are integral.
Combining \lean{trace\_gen\_eq\_sum\_roots} and \lean{trace\_algebra\_map} showed that the trace of $x$ is an integer multiple (namely \lean{finrank F(x) L}) of a sum of conjugate roots, hence we concluded that the trace (and trace form) of an integral element is also integral.

Finally, we showed that the trace form is nondegenerate, following Neu\-kirch~\cite[Proposition~2.8]{Neukirch}.
Since $K / F$ is a finite, separable field extension, it has a power basis \lean{pb} generated by an element $x : K$.
Letting $x_k$ denote the $k$-th conjugate of $x$ in an algebraically closed field $L / K / F$,
the main difficulty was in checking the equality $\sum_k x_k^{i + j} = \Tr_{K / F} (x^{i + j})$.
Directly applying \lean{trace\_gen\_eq\_sum\_roots} was tempting, since we had a sum over conjugates of powers on both sides.
However, the two expressions did not precisely match: the left hand side is a sum of conjugates of $x$, where each conjugate is raised to the power $i + j$,
while the conclusion of \lean{trace\_gen\_eq\_sum\_roots} resulted in a sum over conjugates of $x^{i + j}$.

Instead, the paper proof switched here to an equivalent definition of conjugate:
the conjugates of $x$ in $L$ are the images (counted with multiplicity) of~$x$ under each embedding $\sigma \colon F(x) \to L$ that fixes $F$. This equivalence between the two notions of conjugate was contributed to \mathlib by the Berkeley group in the week before we realized we needed it. Mapping \lean{trace\_gen\_eq\_sum\_roots} through the equivalence gave
$\Tr_{K / F}(x) = \sum_{\sigma} \sigma\ x$.
Since each $\sigma$ is a ring homomorphism, $\sigma\ x^{i + j} = (\sigma\ x)^{i + j}$,
so the conjugates of $x^{i + j}$ are the $(i + j)$-th powers of conjugates of $x$, which concluded the proof.

\section{Class group and class number} \label{sec:class-number}

\subsection{The class group}\label{subsec:class_group}

Recall from Section~\ref{sec math background} that the ideal class group $\Cl_D$ of a Dedekind domain $D$ is the quotient of the group of nonzero fractional ideals of $D$ by the nonzero principal fractional ideals.
More generally, given an integral domain $R$ with fraction field $K$,
we can define the class group $\Cl_R$ as the quotient of the invertible fractional ideals by the nonzero principal fractional ideals.
We formalized this in Lean by first defining a map \lean{to\_principal\_ideal R K $:$ Kˣ → (fractional\_ideal R K)ˣ},
and defined the class group as
\begin{lstlisting}
def class_group (R K : Type*) [comm_ring R] [is_domain R]
  [field K] [algebra R K] [is_fraction_ring R K] :=
(fractional_ideal R K)ˣ ⧸ (range (to_principal_ideal R K))
\end{lstlisting}

Here, \lean{Rˣ} for a semiring $R$ denotes the multiplicative group of its invertible elements.
Recall from Section~\ref{subsection:frac_ideals} that in the general case of an integral domain $R$ 
 the type of fractional ideals of $R$ is endowed with the structure of a commutative semiring.
Therefore, the quotient of the abelian group \lean{(fractional\_ideal R K)ˣ} by the subgroup of nonzero principal fractional ideals is well-defined.
In the case where $R$ is a Dedekind domain, we provided a map \lean{class\_group.mk0} sending nonzero integral ideals of $R$ to the corresponding class in the class group.

\subsection{Finiteness results}\label{subsec:finiteness}

In general, Dedekind domains can have infinite class groups: in fact, a celebrated result by Claborn shows that for every abelian group $G$, there exists a Dedekind domain $D$ with $\Cl_D\cong G$ (see~\cite[Theorem~7]{Cla66}). For an extreme --- but somewhat classical --- example, the Dedekind domain
\[
D=\C[X,Y]/(Y^2-X^3-X-1)  
\]
has class group isomorphic to $\C/\Z^2$. However, as discussed in Section~\ref{sec math background}, the rings of integers of global fields have finite class groups.

We let $K$ be a number field and let $K'$ be a function field, with ring of integers $\OK$ and $\OK[K']$ (we fix a choice of a model $\Fq[q][t]$), respectively. 
Most proofs of the finiteness of $\Cl_{\OK}$ available in a modern textbook (see~\cite[Theorems 4.4,~5.3,~6.3]{Neukirch}) depend on Minkowski's lattice point theorem, a result from the geometry of numbers (which has been formalized in Isabelle/HOL~\cite{Minkowskis_Theorem-AFP}).
Extending this proof to show the finiteness of $\Cl_{\OK[K']}$ is quite involved and does not result in a uniform proof for $\Cl_{\OK}$ and $\Cl_{\OK[K']}$.
Our formalization instead adapted and generalized a classical approach to the finiteness of $\Cl_{\OK}$, where the use of Minkowski's theorem is replaced by the pigeonhole principle. 
We have made available online an informal writeup of the proof, used in the formalization efforts\footnote{\url{https://github.com/lean-forward/class-number-journal/blob/jar-reviews/FiniteClassGroup.pdf}}.
The classical approach seems to go back to Kronecker
and can be found, for instance, in~\cite{Ireland-Rosen}.
We note that some other ``uniform'' approaches can be found in~\cite{Artin-Whaples} and~\cite{Stasinski}.

Let $D$ be an Euclidean domain: in particular, it will be a PID and hence a Dedekind domain. Given a fraction field $F$ of $D$, let $K$ be a finite separable field extension of~$F$.
We formalized, in the theorem \lean{class\_group.fintype\_of\_\-admissible\_\-of\_finite}, that the integral closure $S$ of $D$ in $K$ has a finite class group whenever $D$ has an ``admissible'' absolute value \lean{abs}.
This notion originated in our project from the adaptation and generalization of the classical finiteness proof in interaction with the formalization efforts.
Very informally, the admissibility conditions require that the remainder operator \lean{\%} produces values that are not too far apart.
More precisely, and in more ``ordinary'' mathematical notation, writing $\Mod$ instead of $\%$ and $x \mapsto \lvert x \rvert $ for the absolute value function $D \to \Z$, the latter is called admissible if both:
\begin{itemize}
\item we have a function $\card: \R_{>0} \to \N$;
\item for all $\epsilon \in \R_{>0},\ b \in D-\{0\}$, and finite subsets $A \subset D$,
we can partition $A$
into at most $\card(\epsilon)$ parts, such that all
$x, y \in A$ in the same part satisfy
\[ \lvert x \Mod b - y \Mod b \rvert < \epsilon \lvert b \rvert.\]
\end{itemize}
To formalize this, we made minor modifications like turning $\card$ into a total function on $\R$ and turning $A$ into an $n$-tuple (noting that in this setting there is no need to forbid repetition of elements within the $n$-tuple).
This resulted in the following predicate classifying admissible absolute values \lean{abv}:
\begin{lstlisting}
structure is_admissible (abv : absolute_value D ℤ) extends is_euclidean abv :=
(card : ℝ → ℕ) (exists_partition' :
  ∀ (n : ℕ) {ε : ℝ} (hε : 0 < ε) {b : D} (hb : b ≠ 0)
  (A : fin n → D), ∃ (t : fin n → fin (card ε)),
  ∀ i₀ i₁, t i₀ = t i₁ →
  (abv (A i₁ % b - A i₀ % b) : ℝ) < abv b • ε)
\end{lstlisting}
The \lean{is\_euclidean abv} predicate asserts that the absolute value \lean{abv $:$ D → ℤ} respects the remainder operator of the Euclidean domain $D$, in particular \lean{abv (a \% b) < abv b}.

The above condition formalizes and generalizes an intermediate result in paper proofs of the finiteness of the class group;
the different proofs for number fields and function fields (still assuming $K/F$ separable) become the same after this point.
The direct consequence (by the pigeonhole principle) of admissibility of $x \mapsto \lvert x \rvert $, applied in practice,
is that for all $\epsilon \in \R_{>0},\ b \in D-\{0\},\ n \in \N$, and all subsets $A \in D^n$ containing more than $\card(\epsilon)^n$ elements, there exist distinct $x,y \in D^n$ such that for all $i=1, \ldots, n$ we have $\lvert x_i \Mod b - y_i \Mod  b \rvert <\epsilon  \lvert b \rvert$.
We used division with remainder to replace the \emph{fractional part} operator on $F$ in the classical proof, which was essential to incorporate function fields, and at the same time allowed our proof to stay entirely within $D$ to avoid coercions.

In a similar way to the algebra trace of Section \ref{sec:trace-form}, we defined the norm of an element \lean{x $:$ S} as the determinant of the linear map \lean{lmul x}.
We used the admissibility of \lean{abs} to find a finite set \lean{finset\_approx} of elements of $D$,
such that the following generalization of~\cite[Theorem~12.2.1]{Ireland-Rosen} holds.
\begin{lstlisting}
theorem exists_mem_finset_approx' (a b : S) (hb : b ≠ 0) :
  ∃ (q : S) (r ∈ finset_approx),
  abv (algebra.norm D (r • a - q * b)) < abv (algebra.norm D b)
\end{lstlisting}
Translated back into more ``ordinary'' mathematical notation, this theorem tells us that, for all $a, b \in S$ with $b$ nonzero, there exist $q \in S$ and $r \in \operatorname{finset\_approx}$, such that
\[\lvert \operatorname{Norm}_{S/D}(ra-qb)\rvert < \lvert \operatorname{Norm}_{S/D}(b) \rvert.\]

After this, the classical approach mentioned above formalized smoothly:
we show that each class in $\Cl_{K}$ contains an ideal $J$ with $M \in J$,
where $M$ is the product of all elements of \lean{finset\_approx}, hence $M$ is nonzero.
Since the ideals of the Dedekind domain $S$ have unique factorization,
the nonzero ideal $\langle M \rangle$ spanned by $M$ has only finitely many divisors.
To contain is to divide in Dedekind domains, so there are only finitely many ideals $J$ with $M \in J$.
Thus, we concluded that $\Cl_{K}$ is finite under the condition of the existence of an admissible absolute value on $D$.

It remained to define an admissible absolute value for $\Z$ and $\Fq[q][t]$. On $\Z$, the usual Archimedean absolute value fulfills the requirements
by setting $\card \epsilon$ to be $\frac{1}{\epsilon}$, rounded up.
Since remainders mod $b$ can be chosen to lie in the interval $[0, b[$, partitioning this interval into $\card \epsilon$ intervals of length $\epsilon b$ induces the desired partition.

For $\Fq[q][t]$, we showed that $\lvert f\rvert_{\deg}:=q^{\deg f}$ for $f \in \Fq[q][t]$ is an admissible absolute value.
Fix a polynomial $b \in \Fq[q][t]$ and a set $A' \subset \Fq[q][t]$ of remainders modulo $b$.
Since the coefficients of polynomials in $\Fq[q][t]$ are elements of a finite set of cardinality $q$,
and the degree of each $f \in A'$ is strictly less than $\deg b$,
for each $c$ there are only $q^c$ distinct values for the $c$ coefficients of the monomials of degree $\deg b - c$ up to $\deg b - 1$.
If the highest coefficients of $f, g \in A'$ coincide, then $\lvert (f-g) \rvert_{\deg} < q^{\deg b - c} = q^{-c} \lvert b \rvert_{\deg}$.
By setting $\card \epsilon = \left\lceil{-\log_{q} \epsilon}\right\rceil$ so that $q^{- \card \epsilon} \le \epsilon$,
we can partition $A'$ into $\card \epsilon$ subsets based on highest coefficients,
so that elements of each partition are within distance $\epsilon \lvert b \rvert_{\deg}$ as desired.

We concluded that when $K$ is a global field, restricting to \emph{separable} extensions of $\Fq[q](t)$ in the function field case (but see the remark below), the class group is finite:
\begin{lstlisting}
noncomputable instance : fintype
  (class_group (number_field.ring_of_integers K) K) :=
class_group.fintype_of_admissible_of_finite ℚ K
  absolute_value.abs_is_admissible

noncomputable instance : fintype
  (class_group (function_field.ring_of_integers Fq F) F) :=
class_group.fintype_of_admissible_of_finite (ratfunc Fq) F
  polynomial.card_pow_degree_is_admissible
\end{lstlisting}

Finally, we defined \lean{number\_field.class\_number} and \lean{function\_\-field.\linebreak[0]class\_\-number} as the cardinality of the respective class groups.

We remark that it is possible to get rid of the \lean{[is\_separable F K]} assumption above. 
For instance, using that any function field $K$, given as finite extension of $\Fq[q](t)$, contains an $s \in K$ such that $K/\Fq[q](s)$ is a finite \emph{and separable} extension; see for example \cite[Corollary 4.4 in Chapter VIII]{Lang} (noting that $\Fq$ is perfect and $K$ has transcendence degree $1$ over $\Fq$).
One then also needs to show that the finiteness of the class group of the integral closure of $\Fq[q][s]$ in $K$ is preserved upon replacing $\Fq[q][s]$ by $\Fq[q][t]$.
A trivial way to get rid of the assumption in the statement above is to simply move it to our definition of function field.
While this would be mathematically consistent by the result just cited, we did not opt to do this (for instance showing a finite extension of a function field is a function field
would become nontrivial).
Alternatively, one could aim at dropping the separability condition in the formalized result mentioned in the first paragraph of Section~\ref{sec:integral-closure}.
Having a formalization of this generalization would be interesting in its own right.
This approach would also still need the adaptation of some of the details in the final steps for the finiteness of the class group in the admissible case.

We rounded off our development by determining the class number in the simplest possible case: the rational numbers $\QQ$.
First, we formalized the theorem \lean{class\_number\_eq\_one\_iff}, stating that the class number of $K$ is $1$ if and only if $\OK$ is a principal ideal domain.
After defining the isomorphism \lean{rat.ring\_of\_integers\_equiv} showing $\OK[\QQ]$ is $\Z$,
we could use the fact that $\Z$ is a PID to conclude that the class number of $\QQ$ is equal to $1$:
\begin{lstlisting}
theorem rat.class_number : number_field.class_number ℚ = 1 :=
class_number_eq_one_iff.mpr
  (is_principal_ideal_ring.of_surjective _
    rat.ring_of_integers_equiv.symm.surjective)
\end{lstlisting}

\section{Discussion}

\subsection{Related work}

Broadly speaking, one could see our formalization work as part of number theory. There are several formalization results in this direction.
Most notably, Eberl formalized a substantial part of analytic number theory in Isabelle/HOL~\cite{Eberl19}.
Narrowing somewhat to a more algebraic setting,
Cano, Cohen, Dénès, Mörtberg and Siles formalized in Coq constructive definitions in ring theory, with a particular focus on factorization properties and with applications to algebraic notions like well-founded divisibility and Krull dimension~\cite{linear-algebra-coq}. Moreover, de Lima, Galdino, Borges~Avelar and Ayala-Rincón recently formalized in PVS basic notions regarding ring theory, with a particular focus on quotients: isomorphism theorems, the Chinese remainder theorem and the definitions of prime and maximal ideals~\cite{deLima}. We are not aware of any other formal developments of fractional ideals, Dedekind domains or class groups of rings of integers.

There are many libraries formalizing basic notions of commutative algebra such as field extensions and ideals, including the Mathematical Components library in Coq~\cite{mathcomp},
the algebraic library for Isabelle/HOL~\cite{algebra_isabelle},
the \texttt{set.mm} database for MetaMath~\cite{metamath} and the Mizar Mathematical Library~\cite{algebraic-hierarchy_mizar}.
The field of algebraic numbers, or more generally algebraic closures of arbitrary fields, are also available in many provers.
For example, Blot~\cite{algebraic-numbers-ccorn} formalized algebraic numbers in Coq,
Cohen~\cite{real-algebraic-numbers-coq} constructed the subfield of real algebraic numbers in Coq,
Thiemann, Yamada and Joosten~\cite{algebraic-numbers-isabelle} formalized algebraic numbers in Isabelle/HOL,
Carneiro~\cite{algebraic-numbers-metamath} in MetaMath,
and Watase~\cite{algebraic-numbers-mizar} in Mizar.
To our knowledge, the Coq Mathematical Components library is the only formal development beside ours specifically dealing with number fields~\cite[\texttt{field/algnum.v}]{mathcomp}.

Apart from the general theory of algebraic numbers, there are formalizations of specific rings of integers.
For instance, the Gaussian integers $\Z[i]$ have been formalized
in Isabelle/HOL by Eberl~\cite{gaussian_integers-isabelle},
in MetaMath by Carneiro~\cite{gaussian_integers-metamath}
and in Mizar by Futa, Mizushima, and Okazaki~\cite{gaussian_integers-mizar}.
Eberl's Isabelle/HOL formalization deserves special mention in this context since it introduces techniques from algebraic number theory,
defining the integer-valued norm on $\Z[i]$ and classifying the prime elements of $\Z[i]$.

An application of our work is the formalization of the adèlic ring of a global field in Lean (María Inés de Frutos-Fernández, \cite{Maria-paper}). In particular, the author formalized adic valuations on Dedekind domains, and also proved a correspondence between idèle and ideal class groups. Our work on Dedekind domains and class groups was an essential building block for this project.

Finally, since our project became available in the \mathlib library, a team led by Brasca has begun formalizing Fermat's Last Theorem for regular primes. Fermat's Last Theorem is the assertion that, for all integers~$n\geq 3$,
\begin{equation}\tag*{\refFLT{n}}
\forall x,y,z \in \Z, \quad x^n+y^n=z^n\Longrightarrow x\cdot y \cdot z =0.
\end{equation}
It is immediate to see that, for positive integers $n$ and $m$, if $n$ divides $m$, then the validity of \refFLT{n} implies that of \refFLT{m}.
Therefore, also taking into account that \refFLT{4} was already dealt with by Fermat himself, it suffices to only consider exponents that are odd prime numbers.

Now, an odd prime number $p$ is said to be \emph{regular} if it does not divide the order of the class group $\Cl_{\QQ(\zeta_p)}$ of the number field obtained by adjoining to~$\QQ$ a \emph{primitive $p$th root of unity}~$\zeta_p$; the latter means that~$\zeta_p^p=1$ and~$\zeta_p\neq 1$ or, equivalently, that~$\zeta_p$ is a root of the irreducible polynomial
\[
\frac{X^p-1}{X-1}=X^{p-1}+X^{p-2}+\cdots +X+1.
\]
A classical result, due to Kummer's work in 1847, is that \refFLT{p} is true for every \emph{regular} prime number~$p$. This is the result Brasca and his team have begun formalizing in Lean 3 in the on-going work~\cite{Bra22}, and it evidently requires the finiteness of the class group in order to define the notion of a \emph{regular prime} as above\footnote{It is actually possible to simply \emph{define} a regular prime only in terms of divisibility of some Bernoulli numbers, instead of mentioning class groups. But this definition would at any rate need to be translated in terms of class numbers in order to implement Kummer's proof.}. Moreover, most arguments occurring in Kummer's proof pertain to the structure of the ring of integers $\Z[\zeta_p]=\OK[\QQ(\zeta_p)]$ as a Dedekind domain, and our work lies at the core of the formalization of these structures.

\subsection{Future directions}\label{sec:future_directions}

Having formalized various basic results of algebraic number theory, there are several natural directions for future work, including formalizing some of the following results.
\begin{itemize}

\item The group of units of the ring of integers $\OK^{\times}$ in a number field $K$ is finitely generated, or even Dirichlet's unit theorem~\cite[Theorem 7.4]{Neukirch}, stating that $\OK^{\times}$ has rank $r+s-1$ and that its torsion subgroup is the cyclic group of roots of unity in $K$. Here $r$ denotes the number of real embeddings of $K$ and $s$ the number of conjugate pairs of complex non-real embeddings of $K$. The finite generation result also holds in function fields, again with a precise description of the rank and of the torsion.

\item Other finiteness results in algebraic number theory, most notably Hermite's theorem about the existence of finitely many number fields, up to isomorphism,
with bounded discriminant~\cite[Theorem 2.16]{Neukirch}. While this could be done without interpreting the primes dividing the discriminant as the primes that \emph{ramify} in the number field, it would certainly be interesting to set up some basic ramification theory: on the one hand, this would also prove essential for many other developments and, on the other, it would allow to prove a version of Hermite's theorem stating that, up to isomorphism, there are only finitely many number fields with bounded degree and restricted ramification.
As usual, there are analogous results in the function field setting, though they are less straightforward. One reason for this is that the nondegeneracy of the trace form from Section~\ref{sec:trace-form} does not hold any more when the separability condition is dropped.

\item Class number computations, starting with, say, quadratic number fields.
This could be a step towards the verification of correctness of number-theoretic software, such as KASH/KANT~\cite{kash} and PARI/GP~\cite{PARI2}.
Along the same lines, unit group computations would also be of much interest, most notably the explicit computation of $r+s-1$ generators for the free part of $\OK^{\times}$. 
Restricting to quadratic fields, we see that the rank is positive (and equal to $1$) if and only if the field is of the shape $\QQ(\sqrt{d})$ for some positive integer $d$ that is not a square.
Finding a generator can 
be done by using continued fractions, of which the basics are already implemented in Lean by Kevin Kappelmann, though certifying that a given (perhaps externally computed) element is indeed a generator could also be done without continued fractions.

\item Applications of algebraic number theory to solving Diophantine equations, such as determining all pairs of integers $(x,y)$ such that $y^2=x^3+D$ for
some nonzero $D \in \Z$. It would be interesting to deal with some values of $D$ where no elementary techniques are available and where factorization in the ring of integers of $\QQ(\sqrt{D})$, along with information about the class number, could solve the equation.

\end{itemize}

\subsection{Conclusion}

In this project, we confirmed the rule that the hardest part of formalization is to get the definitions right.
Once this is accomplished, the paper proof (sometimes first adapted with formalization in mind) almost always translates into a formal proof without too much effort.
In particular, we regularly had to invent abstractions to treat instances of the ``same'' situation uniformly.
Instead of fixing a canonical representation, be it $F \subseteq K \subseteq L$ as subfields or the field of fractions $\Frac R$, or the monogenic $K(\alpha)$, we found that making the essence of the situation an explicit parameter, as in \lean{is\_scalar\_tower}, \lean{is\_fraction\_ring} or \lean{power\_basis},
allows to treat equivalent viewpoints uniformly without the need for transferring results.

The formalization efforts described in this paper cannot be cleanly separated from the development of \mathlib as a whole.
The decentralized organization and highly integrated design of \mathlib meant that we could contribute our formalizations as we completed them, resulting in a quick integration into the rest of the library.
Other contributors building on these results often extended them to meet our requirements,
before we could identify that we needed them, as the anecdote in Section \ref{sec:subobjects} illustrates.
In other words, the low barriers for contributions ensured mutually beneficial collaboration.

Quantifying the ratio between the length of our formal proofs and their paper counterparts in an accurate and meaningful way will be very difficult as background assumptions and levels of detail varied significantly. We actually did not always literally follow some written text, but deviated from the paper mathematics (often discussed orally, on blackboards, through Zulip, etc.) on many occasions. An important aspect we had to take into account was to consistently combine different descriptions of mathematical objects from different sources.
The formalization project described in this paper resulted in the contribution of thousands of lines of Lean code involving hundreds of declarations.
A rough estimate concerning the former would be that about five thousand lines of project specific code were added, and about half of that number of lines of more generic background code.
We validated existing design choices used in \mathlib, refactored those that did not scale well
and contributed our own set of designs.
The real achievement was not to complete each proof,
but to build a better foundation for formal mathematics.

\backmatter

\bmhead{Acknowledgements}
We would like to thank Jasmin Blanchette and the anonymous reviewers for useful comments on previous versions of the manuscript, which found their way into this paper.\\
A.~N.~would like to thank Prof.\ Kevin Buzzard for his constant support and encouragement, and for introducing her to the other co-authors.\\
A.~N.~and F.~N.~wish to express their deepest gratitude to Anne Baanen for the generosity shown along all stages of the project. Without Anne's never-ending patience, it would have been impossible for them to contribute to this project, and to overcome several difficulties.\\
Finally, we would like to thank the whole \mathlib community for invaluable advice all along the project.

\bmhead{Declarations}

\paragraph{Funding}
Anne Baanen was funded by NWO Vidi grant No.\ 016.Vidi.189.037, Lean Forward.\\
Sander R. Dahmen was funded by NWO Vidi grant No.\ 639.032.613, New Diophantine Directions.\\
Ashvni Narayanan was funded by EPSRC Grant EP/S021590/1 (UK).

\paragraph{Conflicts of interest}
The authors have no conflicts of interest to declare that are relevant to the content of this article.

\paragraph{Availability of data and material}
See \emph{Code availability}.

\paragraph{Code availability}
Full source code of the formalization is maintained as part of \mathlib, \url{https://github.com/leanprover-community/mathlib}. Copies of the source files relevant to this paper are available in a separate repository at \url{https://github.com/lean-forward/class-number-journal}.

\paragraph{Authors' contributions}
All authors contributed to the formalization project as well as to the extended version. All authors commented on previous versions of the manuscript.

\bibliography{lean}

\end{document}